\documentclass[a4paper]{article}
\usepackage{amsmath,amsfonts,amssymb}
\usepackage{graphicx}
\usepackage{lscape,epsf}
\usepackage{cite}
\usepackage{graphicx}
\usepackage{dcolumn}
\usepackage{bm}
\usepackage{bbm}
\usepackage{amsthm}

\usepackage{dsfont}

\def\ds{\displaystyle}
\def\bea{\begin{array}{c}}
\def\ea{\end{array}}
\def\be{\begin{equation}\bea\ds}
\def\ee{\ea\end{equation}}
\def\bee{\begin{equation}\begin{array}{rcl}\ds}
\def\eee{\end{array}\end{equation}}

\def\d{{\rm d}}

\def\nn{\nonumber}

\begin{document}

\thispagestyle{empty}

\begin{flushright}
ITEP-TH-22/14\\
\end{flushright}

\vspace{30pt}
\begin{center}
{\Large \textbf{Black Holes in AdS/BCFT and Fluid/Gravity Correspondence}}
\end{center}

\vspace{6pt}
\begin{center}
{\large{Javier M.~Mag\'an$^{a}$, Dmitry Melnikov$^{a,b}$ and Madson R.~O.~Silva$^{a,c}$ }\\}
\vspace{25pt}
\textit{\small $^a$  International Institute of Physics, Federal University of Rio Grande do Norte, \\Av. Odilon Gomes de Lima 1722, Capim Macio, Natal-RN  59078-400, Brazil}\\ \vspace{6pt}
\textit{\small $^b$  Institute for Theoretical and Experimental Physics, \\B.~Cheremushkinskaya 25, Moscow 117218, Russia}\\ \vspace{6pt}
\textit{\small $^a$  Departamento de F\'isica Te\'orica e Experimental, Federal University of Rio Grande do Norte, \\ Campus Universit\'ario, Lagoa Nova, Natal-RN  59078-970, Brazil}\\ \vspace{6pt}
\end{center}

\vspace{6pt}

\begin{abstract}
A proposal to describe gravity duals of conformal theories with boundaries (AdS/BCFT correspondence) was put forward by Takayanagi few years ago. However interesting solutions describing field theories at finite temperature and charge density are still lacking. In this paper we describe a class of theories with boundary, which admit black hole type gravity solutions. The theories are specified by stress-energy tensors that reside on the extensions of the boundary to the bulk. From this perspective AdS/BCFT appears analogous to the fluid/gravity correspondence. Among the class of the boundary extensions there is a special (integrable) one, for which the stress-energy tensor is fluid-like. We discuss features of that special solution as well as its thermodynamic properties.
\end{abstract}

\newpage

\section{Introduction}

In physics of real materials edge effects often play an important role. One of the principle examples is the quantum Hall state, for which the sample edges are the loci of all transport in the system, while the bulk of the material is insulating. Another interesting example is the Josephson effect, namely a current across an interface of two superconductors. These phenomena have motivated the authors to study the holographic description of edge or interface physics. (For earlier results in this direction see~\cite{QHall1,QHall2,QHall3,QHall4,Gorsky:2001iq,QHall5,QHall6,QHall7,QHall8,QHall9,QHall10,QHall11,QHall12,QHall13,JJ1,JJ2,JJ3,JJ4,JJ5,JJ6,JJ7,JJ8} and references therein.)

In~\cite{taka} Takayanagi proposed an extension of the AdS/CFT correspondence~\cite{maldacena}, to the case in which the CFT is defined in a space with a boundary, a Boundary Conformal Field Theory (BCFT), \emph{cf.}~\cite{cardy}. The correspondence was dubbed AdS/BCFT. The idea behind the proposal was an appropriate extension of the CFT boundary inside the bulk of the $AdS$ space. The extension (boundary profile in the bulk space, which we henceforth label $Q$) should be dynamical, \emph{e.g.} governed by a variational principle. As proposed in~\cite{taka}, the dynamical feature can be attained via supplementing the variational principle with Neumann boundary conditions on $Q$.

The Neumann boundary conditions used in~\cite{taka} and most of other works imply that the renormalized stress-energy tensor on $Q$ vanishes. This condition is quite restrictive. In particular it appears to be difficult to extend the known solutions of Einstein equations in empty AdS to the case of non-zero temperature. The exercise can be done in the case of $AdS_3$~\cite{taka2,taka3}, but higher dimensional generalizations are still unknown. However, given a boundary condition, one can build a solution perturbatively, as demonstrated in~\cite{taka3}.

In this paper we explore an alternative path. Instead of modifying the bulk geometry we modify the boundary condition itself. This can be done by tuning the energy-momentum tensor residing on $Q$, so as to allow the bulk solution of interest. A modified energy momentum tensor naturally appears when considering some extra degrees of freedom (matter) living on $Q$, which are then expected to be thermally excited by interaction with the black hole thermal radiation. This is very close in spirit to the approach of the fluid/gravity correspondence~\cite{fluidgravity}, which states the equivalence of the certain bulk gravity dynamics and the equations of motion of the dual field theory in the hydrodynamical regime. The connection is clearer if noticed that the gravity definition of the boundary stress-energy tensor, used in fluid/gravity correspondence is equivalent to the Neumann boundary condition for the metric. Therefore one can adopt the fluid/gravity framework in order to study the AdS/BCFT problem.

Within this framework we describe a family of boundary stress-energy tensors $T_{ab}$ residing on $Q$, consistent with the simple $AdS_4$ Schwarzschild black hole in the bulk. Each of the $T_{ab}$ corresponds to a hypersurface in the bulk that bounds a subspace of the black hole solution. We consider the hypersurfaces that preserve all but one spatial translation symmetries. From the point of view of the AdS/BCFT correspondence this is an AdS/CFT problem in a half-space or an infinite strip.

It is well-known in the fluid/gravity correspondence that constant radius slices of the $AdS$ (black hole) space, lead to the stress-energy tensor of a conformal fluid. In this work we are rather interested in those hypersurfaces that extend along the radial direction, from the boundary into the bulk. As a result we find that the corresponding $T_{ab}$ do not generally take the simple fluid-like form. Specifically, in the local rest frame, we find that the stress tensor is not proportional to the unit matrix. This suggests a natural condition to discriminate between different profiles of $Q$, which is to look for energy momentum tensors of the fluid-like form, in local thermal equilibrium with the black hole radiation. We show that there is a unique geometry of the hypersurface that yields a fluid-like $T_{ab}$. Remarkably, the equation for the hypersurface is integrable with the solution given by an elliptic integral.

We further study properties of the special fluid-like solution. The fluid on the boundary extension is subject to a  curved metric, or equivalently to an external field. As a result, the thermodynamic quantities, except entropy density, are coordinate dependent. Nevertheless at every spatial position the fluid has the same equation of state as the well-known conformal fluid defined at the corresponding energy scale. The total entropy of the boundary fluid is consistent with the Bekenstein-Hawking formula. It is proportional to the area of the horizon swept by the boundary hypersurface. Although our work is aiming at the study of edge and interface physics, the solution described in this article might be of interest for the fluid/gravity program itself, as the one exploring all energy scales of the dual field theory.

We also analyze the thermodynamics of the BCFT. The thermodynamics consists of the ``bulk'' and ``boundary'' contributions. The tension of the boundary hypersurface sets a characteristic scale for the corresponding thermodynamical quantities. Computing the boundary contribution to the total action we find that its associated entropy is not equal to the contribution coming from the Bekenstein-Hawking formula alone, as \emph{e.g.} in the 3-dimensional case in~\cite{taka}.

The paper is organized as follows. In section~\ref{sec:AdS-BCFT} we review the AdS/BCFT construction of Takayanagi. In section~\ref{sec:fluid} we describe a family of hypersurfaces and stress-energy tensors consistent with the $AdS_4$ Schwarzschild black hole geometry. We find that the fluid-like condition uniquely defines the geometry of the profile $Q$, and we discuss properties of the boundary fluid. In section~\ref{sec:blackhole} we compute the free energy of the BCFT and derive the boundary contribution to the entropy. We conclude in section~\ref{sec:conclusions}, where we speculate on the dual field theory interpretation and possible applications of our results. In the appendices we review earlier results on AdS/BCFT and fluid/gravity duality. Appendix~\ref{sec:examples} reviews some known solutions of AdS/BCFT, including the BTZ black hole. Appendix~\ref{sec:conformalfluid} overviews the basic example of the conformal fluid in the fluid/gravity framework.


\section{AdS/BCFT}
\label{sec:AdS-BCFT}

The AdS/BCFT correspondence proposed by Takayanagi in \cite{taka} is a suggestion for the gravity dual of a $d$-dimensional CFT defined in a space $M$ with a boundary $P$.\footnote{See the original work of Takayanagi as well as more detailed reviews~\cite{taka2,taka3}. Also see other examples of the AdS/CFT correspondence for the CFT's with boundaries~\cite{AdSCFTb}.} Such CFT's are called boundary CFT's, or BCFT's, if the boundary preserves the $SO(2,d-1)$ subgroup of the $d$-dimensional $SO(2,d)$ conformal group \cite{cardy}.

\begin{figure}
 \centering
 \begin{minipage}{0.45\linewidth}
 \includegraphics[width=\linewidth]{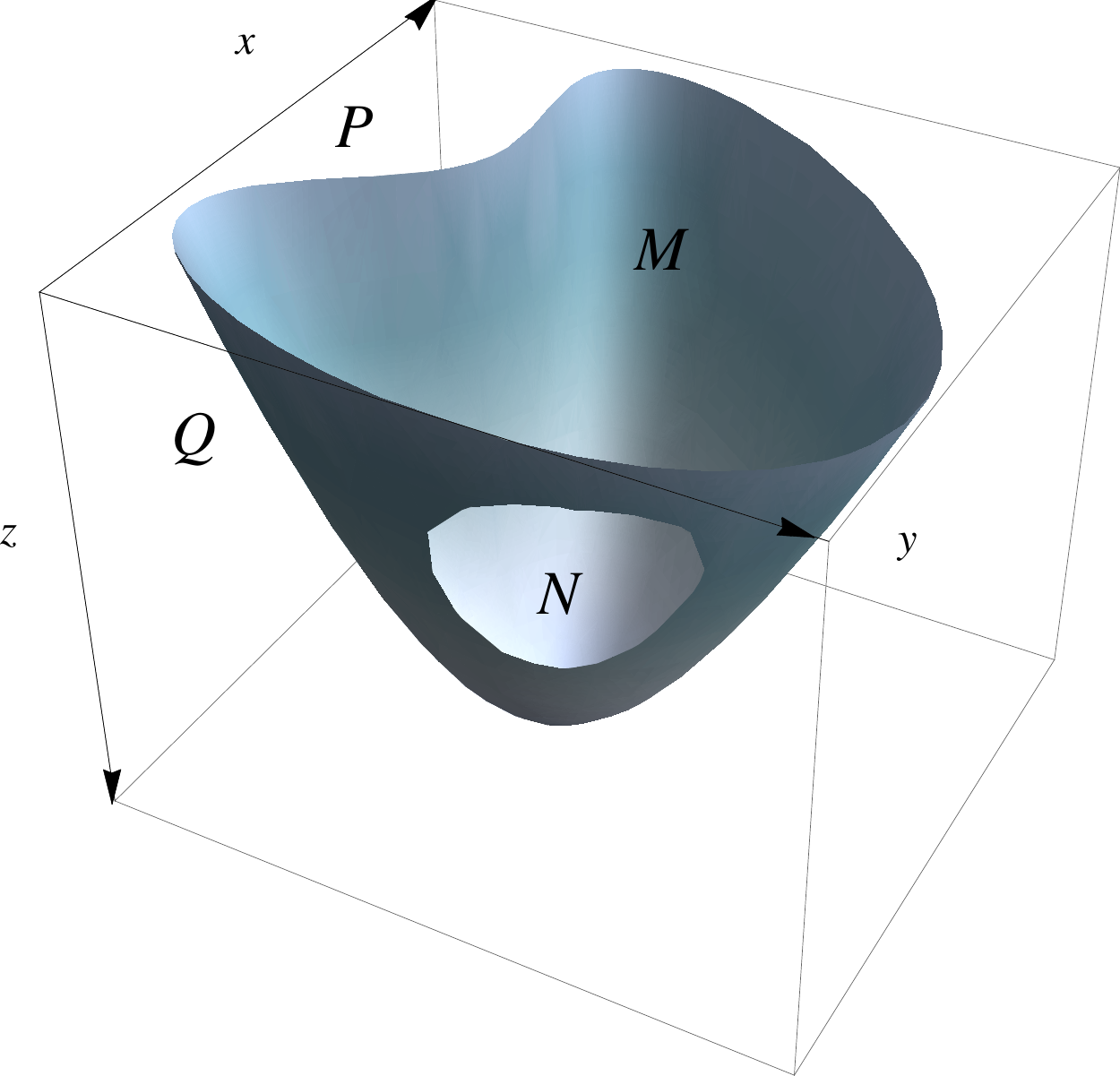}
 \end{minipage}
 \hfill{
\begin{minipage}{0.45\linewidth}
 \includegraphics[width=\linewidth]{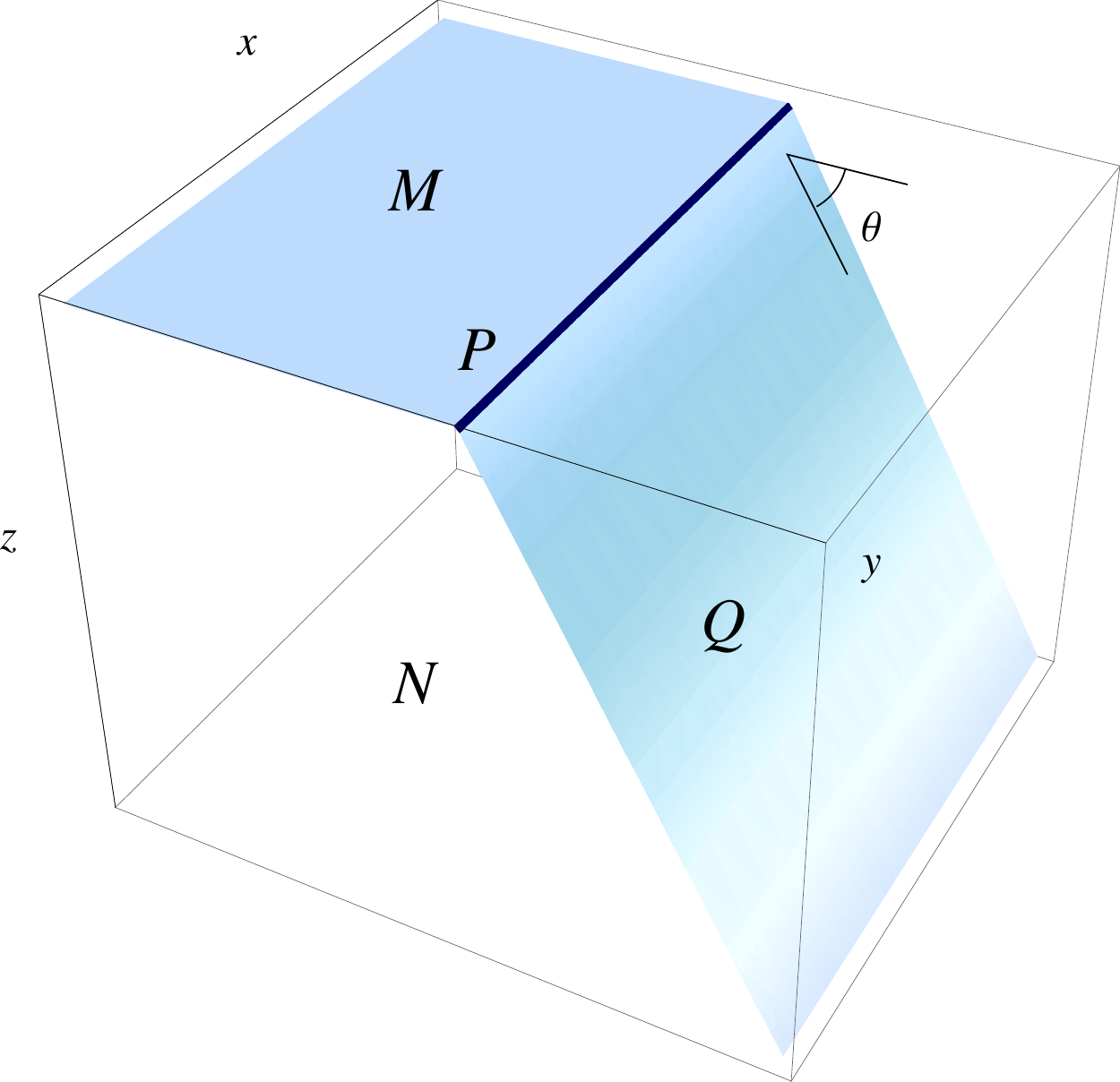}
\end{minipage}
}
\caption{Left: $N$ is the subspace of the bulk of $AdS_{d+1}$, bounded by $Q$. It ``encodes'' physics of $M$. $P$ is the common boundary of $Q$ and $M$. Right: solutions to the AdS/BCFT problem in empty AdS-space are hyperplanes foliating $AdS_{d+1}$ in $AdS_d$ slices.}
\label{fig:AdS/BCFT}
\end{figure}

It is natural to think that the boundary $P$ must be extended to the bulk to cut out a region of the bulk that ``encodes'' the physics of the BCFT. This extension is labelled by $Q$, figure~\ref{fig:AdS/BCFT} (left), while $N$ will label the part of the bulk bounded by $Q$ and $M$: $\partial N=M\cup Q$ and $P=\partial M=\partial Q$. It is also natural to demand that the profile of the boundary $Q$ is determined dynamically. Dynamics can be introduced through the specification of the boundary conditions of the variational problem. For it to be non-trivial one has to choose Neumann boundary conditions for the boundary $Q$ as opposed to the Dirichlet boundary conditions on $M$. Consider the following (4-dimensional) action
\begin{multline}
\label{action}
I =  \frac{1}{2\kappa}\int_N \d^4x\, \sqrt{-g}(R-2\Lambda)+\frac{1}{\kappa}\int_Q \d^3x\, \sqrt{-h}(K-\Sigma) +\frac{1}{\kappa}\int_M\d^3x\, \sqrt{-\gamma}(K^{(\gamma)}-\Sigma^{(\gamma)})+ \\
+ \int_Q\d^3x\, \mathcal{L_{\rm mat}}  + \Delta I\,.
\end{multline}
Here $\kappa=8\pi G$ is the gravitational coupling constant, $g_{\mu\nu}$ is the bulk metric, $h_{ab}$ and $\gamma_{ij}$ are induced metrics on $Q$ and $M$, $K$ and $K^{(\gamma)}$ are corresponding traces of the extrinsic curvature, $\Sigma$ and $\Sigma^{(\gamma)}$ are tensions of $Q$ and $M$ respectively. $\mathcal{L}_{\rm mat}$ is a Lagrangian of possible matter fields on $Q$. $\Delta I$ is the part of the action that contains possible counter-terms and contact terms, localized on $P$. They do not affect the bulk dynamics and will be introduced later.

The variation of action~(\ref{action}) on-shell amounts to the surface terms
\begin{multline}
\delta I = \frac{1}{2\kappa}\int_Q \sqrt{-h}(K_{ab}-\Sigma h_{ab})\delta h^{ab}+\frac{1}{2\kappa}\int_M \sqrt{-\gamma}(K^{(\gamma)}_{ij}-\Sigma^{(\gamma)}\gamma_{ij})\delta\gamma^{ij} -
\\  - \frac{1}{2}\int_Q\d^3x\,\sqrt{-h}T_{ab}\,\delta h^{ab} + \delta(\Delta I)\,,
\end{multline}
where $T_{ab}$ is the matter stress-energy tensor on $Q$. Typically, in the variational principle, one imposes the Dirichlet boundary conditions at the boundary. For the AdS/CFT correspondence this means, in particular, that we fix the boundary metric on $M$, $\delta\gamma_{ij}=0$\,. However one can alternatively choose the Neumann boundary conditions. Let us do this for the induced metric on $Q$, that is let us impose
\be
\label{Neumann-metric}
K_{ab} - (K-\Sigma) h_{ab} = 8\pi G T_{ab}\,.
\ee
This is the dynamical equation for the induced metric $h_{ab}$, or equivalently, for the profile of the hypersurface $Q$. The solution of the AdS/BCFT then requires a solution of the equations of motion, derived from action~(\ref{action}), which additionally satisfies the Dirichlet boundary conditions on $M$ and Neumann boundary conditions~(\ref{Neumann-metric}) on $Q$.\footnote{Although the AdS/BCFT is formulated in a bottom-up fashion one can embed it in a top-down string theory construction~\cite{taka2}.}

In this work we are interested in studying the holographic models of boundary CFT's at finite temperature. For this we would like to find a solution of the AdS/BCFT problem, which has an asymptotically $AdS_4$  black hole geometry. For simplicity, we will consider the boundary $P$ specified by the condition $y={\rm const}$, where $y$ is one of the coordinates on $M$, figure~\ref{fig:AdS/BCFT} (right), that is we consider the AdS/BCFT problem on a half of Minkowski space.

To solve equation~(\ref{Neumann-metric}) one needs to specify an energy-momentum tensor $T_{ab}$ for the matter fields on $Q$. In the simplest case~\cite{taka} one assumes $T_{ab}=0$.\footnote{Notice that here, as opposed to~\cite{taka}, we do not consider the vacuum term $-\Sigma h_{ab}/{8\pi G}$ to be part of $T_{ab}$.} With this assumption one can solve the problem for the empty AdS case, and for the BTZ black hole in three dimensions. These two cases are reviewed in detail in appendix~\ref{sec:examples}. On the other hand it is easily checked that the AdS-Schwarzschild (or other plane-symmetric) solution does not satisfy~(\ref{Neumann-metric}) with $T_{ab}=0$ for a number of dimensions greater than three, except for $\Sigma=0$. One possibility, explored in~\cite{taka3}, is to construct a solution as a perturbative expansion over the  vacuum AdS solution. This has however the disadvantage of working with complicated metric from the start.

Another possibility, which we explore in this article, is to allow matter on $Q$. Specifically, the objective of this work will be to analyze possible theories on $Q$, which would be consistent with the Neumann boundary conditions and the $AdS_4$-Schwarzschild black hole geometry:
\be
\label{ansatz-metric}
\d s^2 = \frac{L^2}{z^2}\left(- f(z)\d t^2 + \d x^2 + \d y^2 + \frac{\d z^2}{f(z)} \right),
\ee
where the function $f(z)$ is given by
\be
f(z)=1-\left(\frac{z}{z_h}\right)^3 \,.
\ee
This solution describes a black hole with the horizon radius $z_h$ and the Hawking temperature
\be
\label{HawkingT}
T_H=\frac{3}{4\pi z_h}\,.
\ee


\section{Boundary fluid from AdS/BCFT}
\label{sec:fluid}

As explained in the previous section, there are two approaches to the AdS/BCFT problem. In the first, one fixes the boundary stress (-energy-momentum) tensor, \emph{e.g.} $T_{ab}=0$, or in a more general form, and finds a solution to the bulk Einstein equations, satisfying~(\ref{Neumann-metric}). In the other, one fixes the bulk metric and finds an appropriate matter content on $Q$, parameterized by stress-energy tensors $T_{ab}$, which supports the metric of interest. Put this way the AdS/BCFT problem is in fact equivalent to a holographic duality, between gravity in the bulk and the matter theory on the boundary $Q$, analogous to the AdS/CFT correspondence itself.\footnote{AdS/CFT correspondence with Neumann boundary conditions was previously discussed in~\cite{Compere:2008us}. AdS/BCFT may be considered as an extension of that approach to the case of arbitrary boundaries.} In a particular regime it is equivalent to the fluid/gravity correspondence~\cite{fluidgravity}, which we shall discuss momentarily.

The fluid/gravity correspondence is based on the observed equivalence of the Einstein equations with a negative cosmological constant in the bulk of a space-time and fluid-like equations on a time-like hypersurface, which is considered as a boundary of that space-time. The equations of the fluid are nothing but the statement of the covariant conservation of the stress-energy tensor seen by an observer placed on the hypersurface. In the long-wave hydrodynamical regime this is equivalent to the equations of relativistic hydrodynamics. The correspondence is a duality of two different descriptions of the same physics. By duality, one can reconstruct the bulk metric from a given solution to the hydrodynamical equations and vice-versa.

Given a hypersurface $Q$ the stress-energy tensor $T_{ab}$ residing on it is defined through the variation of the action with respect to the induced metric on $Q$. Naive unrenormalized action may lead to a diverging $T_{ab}$, or physical quantities computed from it. The divergences can be cancelled by subtracting an appropriately chosen vacuum solution (AdS in our case), or equivalently by adding appropriate boundary terms as specified by the procedure of holographic renormalization~\cite{skenderis,vijay}. More specifically the renormalization procedure leads to the following form of $T_{ab}$:
\be
\label{Tab}
T_{ab} = -\frac{L^{d-2}}{\kappa z^{d-2}}\left(K_{ab}-K h_{ab}+\Sigma h_{ab} -\kappa T_{ab}^{(R)} - \kappa T_{ab}^{(ct)}\right).
\ee
Notice that this is the ``intrinsic'' stress-energy tensor, which is defined with respect to the intrinsic hypersurface metric,
\be
\label{intr-metric}
\hat{h}_{ab}=h_{ab}\cdot \frac{z^2}{L^2}\,.
\ee
In the expression above we have written explicitly the part coming from the Gibbons-Hawking and tension-like terms in~(\ref{action}), while $T_{ab}^{(R)}$ and $T_{ab}^{(ct)}$ are additional possible contributions from the intrinsic curvature and counter-terms respectively.

As follows from the discussion, the fluid/gravity correspondence, within the limit of its validity, is equivalent to AdS/BCFT. The Brown-York type procedure of extracting the dual stress-energy tensor~(\ref{Tab}) can be mathematically formalized as the variational principle for~(\ref{action}) supplemented with Neumann boundary conditions~(\ref{Neumann-metric}). The $\mathcal{L}_{\rm mat}$-term in the action~(\ref{action}) plays a role of a source term with respect to the boundary (bulk) metric. In what follows we identify the stress-energy tensor of the matter fields with the $T_{ab}$ in the left hand side of~(\ref{Tab}). From this perspective, finding a finite temperature solution to the AdS/BCFT problem~(\ref{Neumann-metric}) translates into finding a dual fluid theory living on $Q$ supporting the black hole solution~(\ref{ansatz-metric}).

Considering the problem in a half-space $y<0$, let us parameterize a generic hypersurface $Q$ by the function $y(z)$, figure~\ref{fig:AdS/BCFT} (right), and restrict to the first three terms in~(\ref{Tab}). We then find the boundary stress-energy tensor in the following form
\be
\label{Tab-gen}
T_{ab} = \left(
\begin{array}{ccc}
- \hat{h}_{tt}\, \epsilon(z) && \\
& \hat{h}_{xx}p_x(z) & \\
&&  \hat{h}_{zz} p_z(z)
\end{array}
\right),
\ee
where $\hat{h}_{ab}$ is the intrinsic metric~(\ref{intr-metric}) on $Q$ and the following functions have been introduced
\begin{eqnarray}
\label{epsilon}\epsilon(z) & = & \frac{L^2}{2\kappa z^3}\left(2\Sigma L+\frac{(zf'-4f)y'(z)-4f^2y'(z)^3+2zfy''(z)}{(1+fy'(z)^2)^{3/2}}\right) ,\\
p_x(z) & = & \frac{L^2}{2\kappa z^3}\left(-2\Sigma L+\frac{2(2f-zf')y'(z)+f(4f-zf')y'(z)^3-2zfy''(z)}{(1+fy'(z)^2)^{3/2}}\right), \\
p_z(z) & = & \frac{L^2}{2\kappa z^3}\left(-2\Sigma L+\frac{y'(z)(4 f-z f')}{\sqrt{1+fy'(z)^2}}\right).
\end{eqnarray}

Close to the boundary $z\to 0$ the bulk metric approaches that of the empty $AdS$-space. Therefore we expect to recover solution~(\ref{QprofileT}) of appendix~\ref{sec:emptyAdS} in this limit. For this reason we may set $\Sigma L =2\cos\theta$, where $\theta$ is the angle that the profile $Q$ makes with the $y$-axis, figure~\ref{fig:AdS/BCFT} (right). With this condition~(\ref{Tab-gen}) describes a generic theory on the boundary hypersurface $Q$, which supports the Schwarzschild black-hole solution in the bulk.

\begin{figure}
 \begin{minipage}{0.45\linewidth}
 \includegraphics[width=\linewidth]{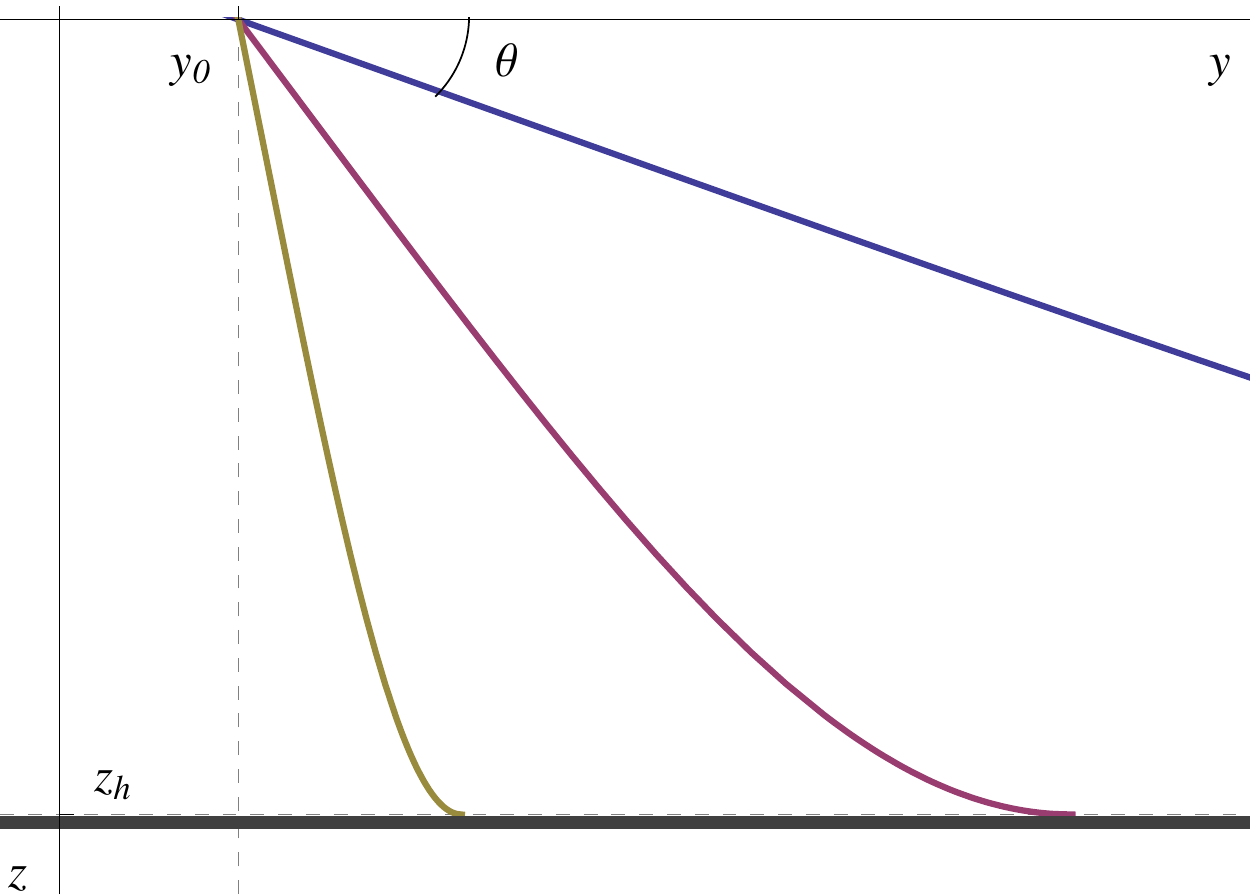}
 \end{minipage}
 \hfill{
\begin{minipage}{0.45\linewidth}
 \includegraphics[width=\linewidth]{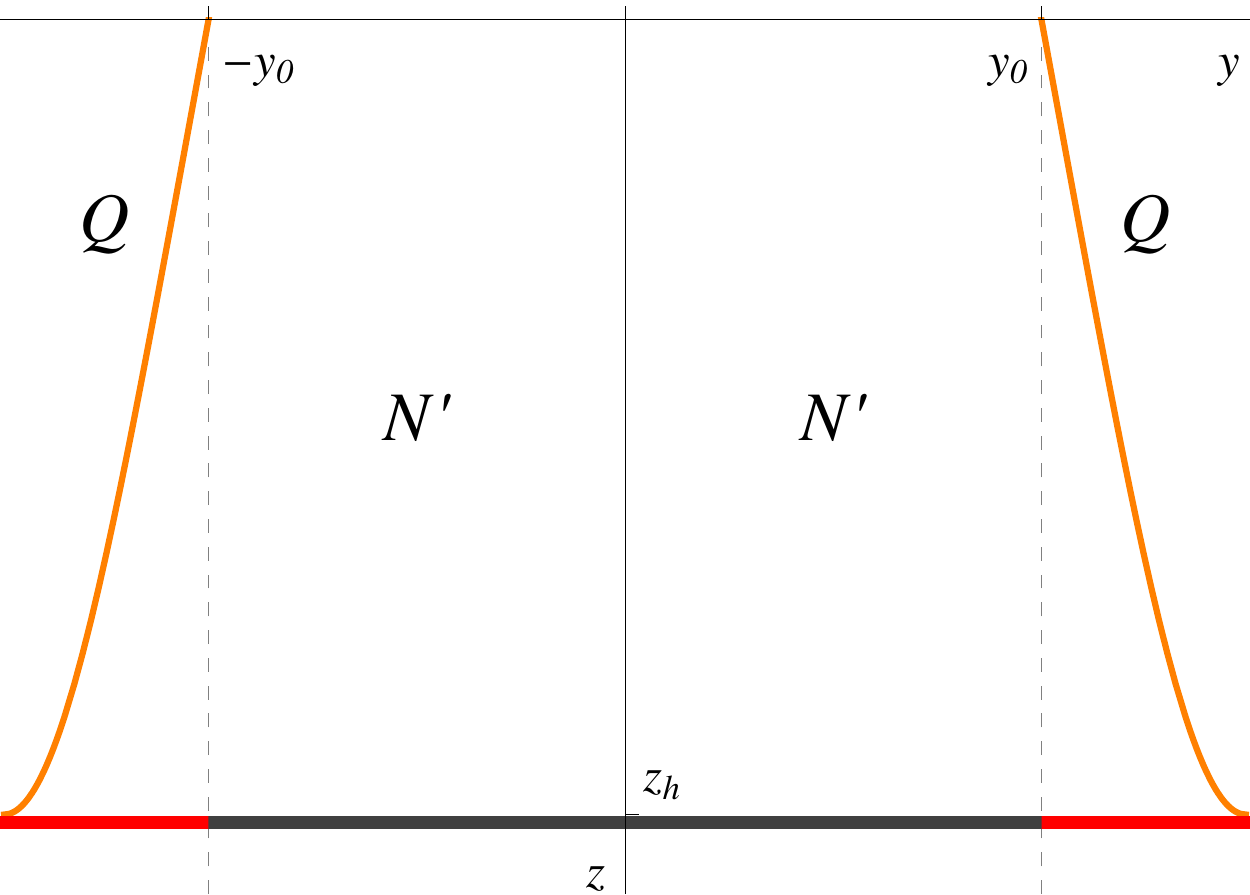}
\end{minipage}
}

 \caption{Profiles of the brane $Q$ described by (\ref{profileQ1}) for different angles $\theta$ (left). AdS/BCFT model of an infinite stripe (right). $N'$ is a part of the bulk bounded by the dashed lines. Areas of the horizon ``swept" by the branches of the hypersurface $Q$ are colored with red.}
 \label{fig_profile1}
\end{figure}

Interestingly generic $T_{ab}$ does not describe a fluid-like system ($p_x\neq p_z$). One can ask whether the above class of stress-energy tensors can at all describe a fluid. Apparently this is a restriction on the profile $y(x)$. Demanding $p_x=p_z$ one arrives at a particularly simple equation
\be
\label{Yeqn0}
 \frac{2f y''(z) + f'y'(z)}{\sqrt{1+fy'(z)^2}} =0\,,
\ee
which can be easily integrated to get
\be
\label{Yeqn1}
f{y'}^2 = {\rm const}\,.
\ee
So, the general solution, which yields a fluid-like theory on $Q$, is provided by the profile
\be
\label{profileQ1}
y=y_0 + \int\limits_0^z \frac{\cot\theta\d q}{\sqrt{f(q)}}\,,
\ee
parameterized by the ``contact" angle $\theta$. For $f=1$ the result~(\ref{QprofileT}) of Takayanagi is reproduced, while for the black hole solution $f=1-z^3/z_h^3$ the profiles are shown on figure~\ref{fig_profile1} (left) for different values of $\theta$. In the limit $\theta\to 0$ we recover the well-known results of the conformal fluid (see appendix~\ref{sec:conformalfluid}). In particular, $p_x=p_z=p$ in this limit.

More interesting is the behaviour close to the horizon. The profile is regular and has the expansion
\be
y= {\rm const} + \frac{2\cot\theta}{\sqrt{3}}\sqrt{1-z}+ O\left((1-z)^{3/2}\right)\,,
\ee
that is it reaches the horizon at a finite value $y_h$. This value gives a characteristic distance scale $\Delta y=y_h-y_0$ set by angle $\theta$. The solution can be extended by adding a second branch of the surface $Q$ as shown in figure~\ref{fig_profile1} (right). This way one gets a configuration, similar to the one found in the BTZ case~\cite{taka}, \emph{cf.} figure~\ref{fig:BTZ} from appendix~\ref{sec:BTZ}. Notice that the walls of the configuration in figure~\ref{fig_profile1} (right) must satisfy $y'(z)>0$ ($y'(z)<0$) for the right (left) branch of the boundary $Q$. Otherwise one will end up with a theory with negative energy density and temperature (null/weak energy condition).

\begin{figure}[htb]
 \begin{minipage}{0.45\linewidth}
 \includegraphics[width=\linewidth]{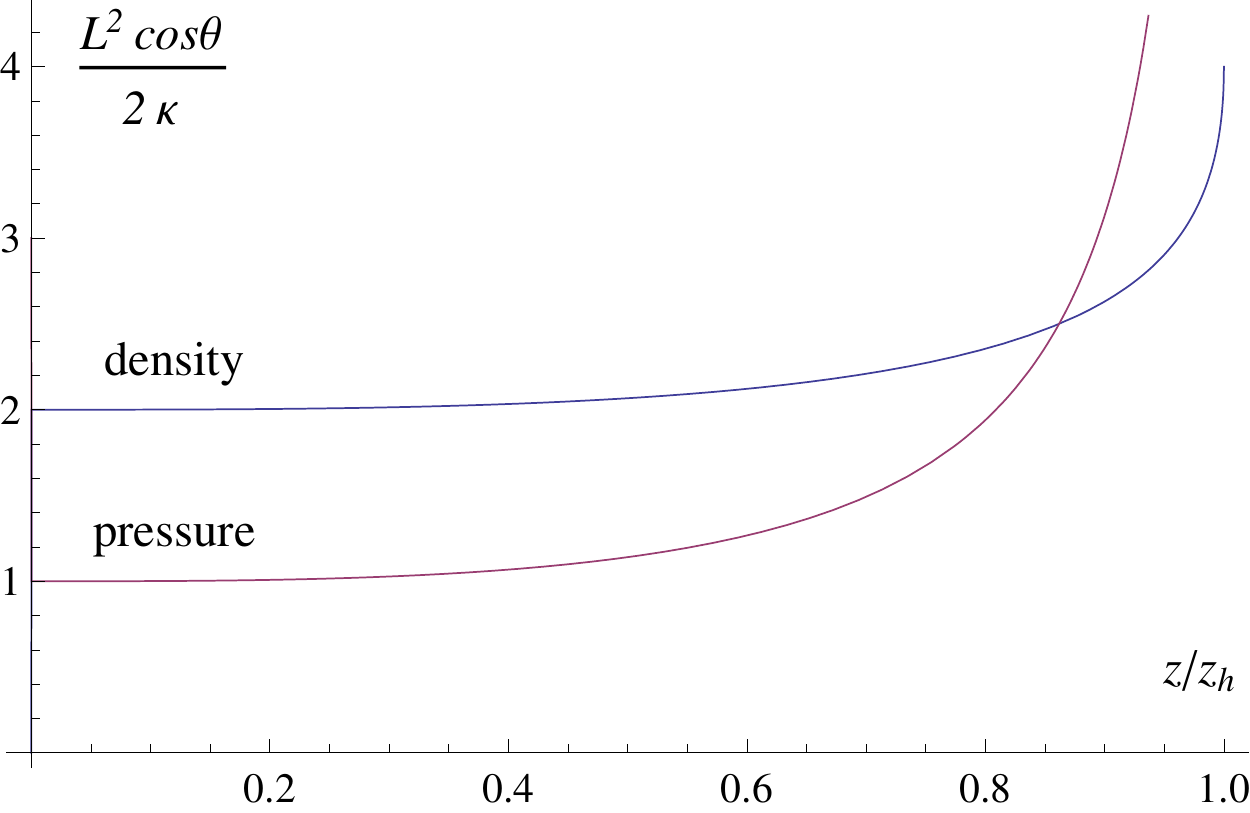}
 \end{minipage}
 \hfill{
\begin{minipage}{0.45\linewidth}
 \includegraphics[width=\linewidth]{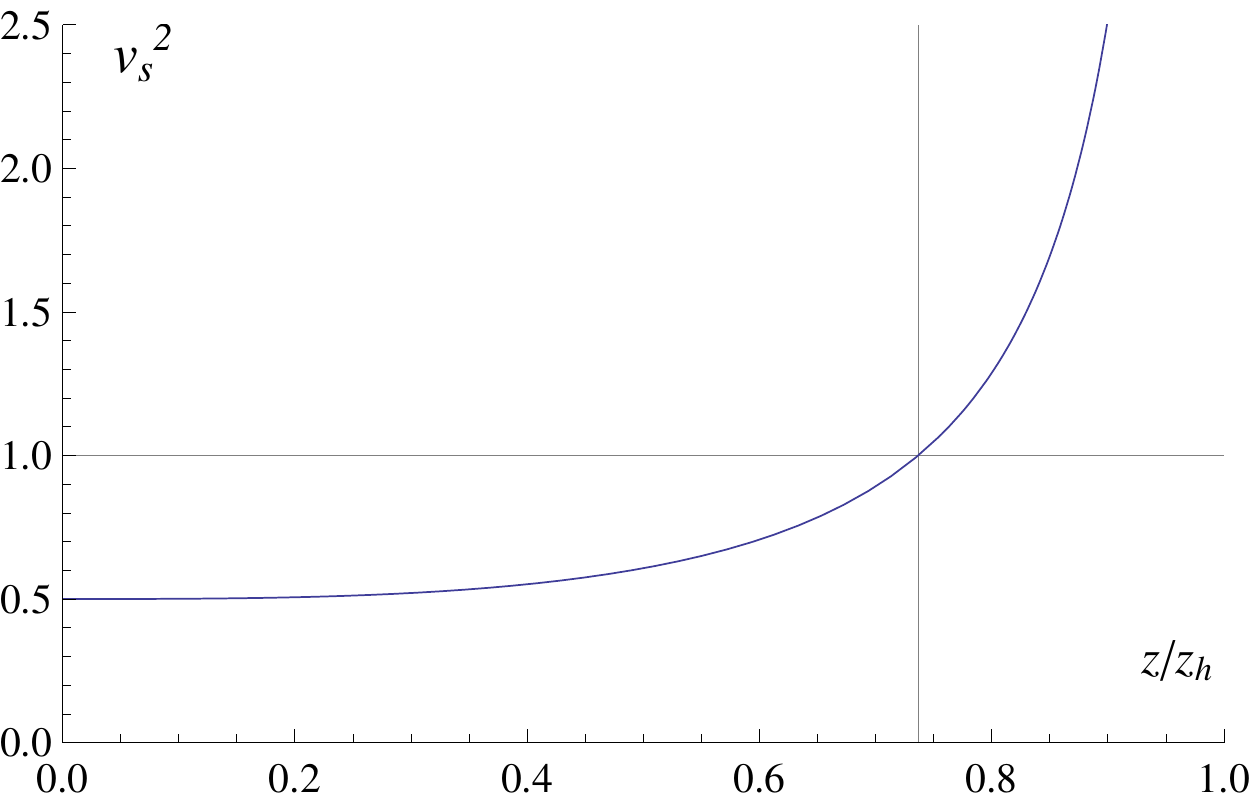}
\end{minipage}
}
\caption{Renormalized energy density (blue) and pressure (magenta) in the effective theory on the surface $Q$ (left). Dependence of the square of speed of sound on the position on $Q$ (right).}
\label{fig:density-pressure}
 \end{figure}

The renormalized energy density and pressure on the hypersurface $Q$ are given by
\begin{eqnarray}
\label{edensity} \epsilon & = &  \frac{2L^2\cos\theta}{\kappa z^3}\left(1-\sqrt{f}\right), \\
\label{pressure} p & = &  \frac{L^2\cos\theta}{2\kappa z^3\sqrt{f}}\left(4f-zf'-4\sqrt{f}\right).
\end{eqnarray}
For a given value of the coordinate $z$ these are the results~(\ref{ezconst}) and~(\ref{pzconst}) for the conformal fluid renormalized by a factor $\cos\theta$, see appendix~\ref{sec:conformalfluid}. Put differently, the two fluids have the same equation of state. The dependence of $\epsilon$ and $p$ on the position on $Q$ (equivalently on coordinate $y$) is demonstrated on figure~\ref{fig:density-pressure} (left). 

As hypersurface $Q$ has a non-flat induced metric, the fluid is subject to a gravitational force, or, in other words, it is subject to an external field. As a result equilibrium thermodynamical quantities depend on the coordinates. The dependence on the $z$-coordinate is interpreted as a renormalization of physical quantities. In this respect our fluid is made of ``dressed" filaments of the same fluid~\cite{Brattan:2011my}. Close to the boundary the system is conformal $\epsilon=2p$. At the opposite end the system is strongly non-conformal: while the energy density is finite, the pressure diverges as $y\to y(z_h)$. Similar behavior is well known in the models of fluid dynamics~\cite{Bludman:1968zz}, see~\cite{Brattan:2011my} for the discussion in the context of fluid/gravity correspondence.

Assuming the local temperature on the surface $Q$ is given by
\be
\label{localT}
T=\frac{T_H}{\sqrt{f(z)}}\,,
\ee
the local entropy density is
\be
\label{Yentropy}
s=\,\frac{\epsilon+p}{T}= \frac{L^2\cos\theta}{4Gz_h^2}\,.
\ee
We notice that the entropy density is constant over the surface. This is consistent with the general observation that the entropy of the fluid on a hypersurface in the bulk is proportional to the area of the horizon swept by the hypersurface. Indeed, if this entropy density is integrated with respect to the intrinsic metric~(\ref{intr-metric}) on $Q$, we find that the total entropy is

\be
\label{Sfluid}
S_Q= \frac{L^2(y(z_h)-y_0)}{4Gz_h^2}\,\Delta x\,
\ee
for each branch of the configuration, consistent with the Bekenstein-Hawking entropy.

Although we have concentrated in four space-time dimensions the results of this section can be easily generalized to the case of the $AdS_{d+1}$-Schwarzschild black hole with $d>3$. There is a unique hypersurface $Q$ satisfying the condition $p_z=p_i$, where $i$ label the spatial coordinates transverse to $y$. The profile of the hypersurface stays defined by equation~(\ref{Yeqn1}) with function $f(z)$ appropriately modified.


\section{Black hole thermodynamics}
\label{sec:blackhole}

Let us derive the equilibrium thermodynamics of the BCFT from the gravity configuration described by figure~\ref{fig_profile1} (right). The free energy of the system is computed from the total Euclidean action. We split the latter in two parts: one, which we call the ``bulk'' contribution, will be independent from the tension parameter (angle $\theta$); and the other, ``boundary" contribution, will be $\theta$-dependent. The bulk part of the total (renormalized) action reads
\be
\label{Sbulk}
I_{\rm bulk}= -\frac{1}{16\pi G}\int_{N'} \d^4x\, \sqrt{g_E}(R-2\Lambda) - \frac{1}{8\pi G}\int_M \sqrt{\gamma}(K^{(\gamma)}-\Sigma^{(\gamma)})= -\frac{L^2}{12Gz_h^2}\,V\,,
\ee
where the first integral is taken over the subspace $N'$ of the bulk space $N$, which excludes the two wedges underneath the hypersurface $Q$, figure~\ref{fig_profile1} (right). $V$ stands for the volume of the BCFT. In this calculation the tension of the boundary $M$ must be selected to be $\Sigma^{(\gamma)}=2/L$. The renormalization procedure here is equivalent to subtracting the action of the vacuum AdS solution. In both ways one derives the standard AdS/CFT result.

The $\theta$-dependent part consists of two pieces: the contribution from the bulk of the wedges\footnote{In this calculation one may find useful formulae for the integrals presented in appendix~\ref{sec:appendix}.}
\begin{multline}
\label{SbulkQ}
2\times\frac{6\beta L^2}{16\pi G }\int \d x\, \int\limits_{\epsilon}^{z_h} \frac{\d z}{z^4}\int\limits_{y_0}^{y(z)} \d y = \frac{z_h L^2}{G } \,\cot\theta\int \d x\, \int\limits_{\epsilon}^{z_h} \frac{\d z}{z^4}\int\limits_0^z\frac{\d q}{\sqrt{f(q)}} =
\\ = -\frac{z_h L^2}{3G } \, \Delta x \left(\frac{y(z_h)-y_0}{z_h^3}-\frac{y(\epsilon)-y_0}{\epsilon^3}-\frac{y(z_h)-y_0}{4z_h^3}-\frac{\cot\theta}{2\epsilon^2}+O(\epsilon)\right)
\end{multline}
and the boundary action part
\begin{multline}
\label{SQ}
2\times -\frac{1}{8\pi G}\int \d^3x\, \sqrt{h_E}(K-\Sigma) = \frac{z_hL^2}{6G}\, \cot\theta\int\d x \int\limits_\epsilon^{z_h}\frac{\d z}{z^3}\,\left(4+\frac{zf'(z)-6f}{\sqrt{f(z)}}\right)=
\\ = -\frac{L^2\Delta x}{3Gz_h}\,\cot\theta\left(1-\frac{z_h^2}{\epsilon^2}\right) - \frac{L^2\Delta x}{2Gz_h^2}\left(y(z_h)-y(\epsilon)\right) - \frac{z_hL^2}{G}\,\Delta x\left(-\frac{3(y(z_h)-y_0)}{4z_h^3}+\frac{\cot\theta}{2\epsilon^2}+O(\epsilon)\right).
\end{multline}

The bulk action possessed cubic divergences, which were cancelled by the Gibbons-Hawking term on the boundary $M$. The boundary contribution has quadratic divergences. Adding~(\ref{SbulkQ}) and~(\ref{SQ}) do not remove the divergence completely, only up to the $(y(\epsilon)-y_0)/\epsilon^3$ term. The remaining divergence is the UV divergence of the Takayanagi's vacuum solution~(\ref{QprofileT}) reviewed in appendix~\ref{sec:emptyAdS}. We also observe a similar (linearly-divergent) term in the lower-dimensional example, as reviewed in the appendix~\ref{sec:BTZ}, equation~(\ref{Ibdry1}).

The remaining divergence can be removed by a counter-term at the boundary $P=Q\cap M$ of the CFT:
\be
-\,\frac{1}{4\pi G}\int_P\d^2x\,\sqrt{\sigma}\cot\theta\,,
\ee
where $\sigma_{ij}$ is the induced metric on $P$. Adding extra counter-terms on $P$ does not affect the dynamics discussed before.\footnote{This counter-term would be taken into account automatically should one renormalize the action via the subtraction of the vacuum (empty AdS) solution.} One should also remember that since the boundary $P$ between $Q$ and $M$ is not smooth one has to add an extra counter-term
\be
\,\frac{1}{8\pi G}\int_P\d^2x\,\sqrt{\sigma}\,\theta\,
\ee
to cancel the divergency associated with the discontinuity of the normal vector across $P$~\cite{Hayward:1993my}. Indeed, the extrinsic curvature $K$ depends on the second derivative of the profile $y(z)$. Since the profile function is not smooth, the second derivative contains a delta-function with support on $P$. The above (infinite) counter-term merely cancels the contribution of this delta-function. The importance of the second counter-term was stressed in~\cite{taka3}. Specifically the stress-energy tensor defined on $P$ needs this term to correctly compute the Weyl anomaly in the AdS$_3$/CFT$_2$ case.

After all the divergencies are taken care of one arrives at
\be
\label{Sbdry}
2I_{\rm bdry}= - \frac{L^2\cot\theta}{3Gz_h}\,\Delta x \,.
\ee
Now the free energy of the configuration illustrated by figure~\ref{fig_profile1} (right) reads
\be
\label{freee}
F=T_HI_E = T_HI_{\rm bulk} +2T_HI_{\rm bdry}= -\frac{L^2V}{16\pi G z_h^3} - \frac{L^2\cot\theta}{4\pi G z_h^2}\,\Delta x\,.
\ee
From the free energy one can compute the entropy
\be
\label{Snew}
S\equiv S_{\rm bulk} + 2S_{\rm bdry} = -\frac{\partial F}{\partial T_H}= \frac{L^2V}{4Gz_h^2}+ 2\,\frac{L^2\cot\theta}{3G z_h}\,\Delta x\,,
\ee
and the internal energy of the configuration,
\be
E\equiv E_{\rm bulk} + 2E_{\rm bdry}=\frac{\partial I}{\partial \beta}= \frac{L^2V}{8\pi G z_h^3} + \frac{L^2\cot\theta}{4\pi G z_h^2}\,\Delta x\,.
\ee
The above expressions satisfy the thermodynamical relation $F=E-T_HS$.

We note that as far as the boundary contribution is concerned the above entropy is not the one of the Bekenstein-Hawking formula~(\ref{Sfluid}). While it is the case for the AdS/BCFT of the BTZ black hole, appendix~\ref{sec:BTZ}, here we find that $S_Q\simeq 1.05S_{\rm bdry}$. We are not aware of the reason, why the two entropies should be the same, or approximately the same. It would be interesting to derive the relation between them. For example, one can notice that the above boundary free energy can be calculated from the fluid data as follows
\be
2F_{\rm bdry}= - \int_Q p\sqrt{f}\,\d V\,,
\ee
where the volume integration should be performed with respect to the intrinsic metric~(\ref{intr-metric}).


\section{Conclusions}
\label{sec:conclusions}

In this work we have studied the AdS/BCFT problem at finite temperature. We looked for the solutions of the AdS/BCFT boundary conditions consistent with the planar $AdS$-Schwarzschild geometry in the bulk. Such solutions can be found if certain matter field fields are introduced on the AdS/BCFT boundary $Q$. The solutions can be classified by the stress-energy tensor of the matter theory. As the main result of our investigation we have noticed that among a continuum of the profiles of $Q$ (or stress-energy tensors $T_{ab}$) there is a single profile, for which gravity yields a fluid-like stress-energy tensor, in local thermodynamic equilibrium with the black hole radiation. The profile function in this case happens to be integrable, provided by an elliptic integral.

From the perspective of this special solution, the AdS/BCFT problem turns out to be equivalent to the problem of the fluid/gravity correspondence. Indeed, one can understand the fluid/gravity correspondence as a problem of finding a dual description of the bulk gravity theory in terms of a given fluid living on a boundary of the bulk space. The stress-energy tensor of the fluid in this case acts as a source for the boundary metric.

Having noticed the equivalence with fluid-gravity correspondence we have analyzed the matter theory on $Q$ with ideal-fluid-like $T_{ab}$. We have found that this special solution yields the same ``conformal'' fluid of the standard fluid/gravity calculation in the $AdS$-Schwarzschild geometry, albeit with thermodynamical quantities dependent on position (inhomogeneous fluid). We have computed the local quantities describing this boundary fluid and shown that they satisfy the first law of thermodynamics, describing a fluid in local thermal equilibrium with the Hawking radiation of the black hole. In particular, we have found that the entropy density does not depend on the coordinates and is consistent with the Bekenstein-Hawking formula, which tells that the total entropy is proportional to the area of the horizon swept by the hypersurface $Q$.

The coordinate dependence reflects the fact the fluid is subject to an external force with thermodynamical quantities renormalized according to the position. Our fluid explores then all energy scales of the boundary field theory, in particular that of the boundary edge $P$. This is a novel feature which we have not found in the literature and that we would like to understand better in future work. It might be interesting for the fluid/gravity program itself.

Finally we have computed the free energy of the BCFT, its entropy and internal energy. They satisfy the thermodynamic relation $F=E-TS$. Interestingly, although the boundary entropy does not exactly coincide with the Bekenstein-Hawking entropy, proportional to the horizon area swept by the profile $Q$, the difference between them is rather small. We remark that there is no \emph{a priori} reason for this to be the case, as the boundary entropy of $P$ is not directly related with the black hole entropy. The latter correctly reproduce the entropy of the fluid living on $Q$. It is nevertheless interesting to better understand the relation between the two.

We believe that the most interesting direction to develop from our results is the extension of this AdS/BCFT framework to a model of interface physics\footnote{Also see the recent developments on the holographic constructions of boundary/defect CFTs~\cite{Defects1,Defects2,Defects3,Defects4,Defects5,Defects6}.}. Indeed in the Takayanagi's construction there is no part of the bulk space beyond $N$, the part of interest. It seems quite natural to extend this correspondence to an interface problem, where the bulk space exists on both sides of the boundary $Q$, and the interface separates two distinct phases, such as holographic superconductors~\cite{Gubser:2008px,Hartnoll:2008vx} with different order parameters, simulating a holographic Josephson junction, or phases at different filling fraction, in the quantum Hall effect, \emph{cf.}~\cite{QHall7,QHall9,Defects4}. Indeed, an interesting feature of the solution concerning these problems is that the geometry of the profile $Q$ defines a finite ``width'' associated with the interface $P$. We expect this width to be related to some potential barrier of the boundary interface, such as a penetration length or escape energy from $P$. The volume associated to this width become parameter-dependent, which indicates that such a system is a candidate for a model of a gapped system with a gapless boundary.

\vspace{0.5cm}
\paragraph{Acknowledgements.} The authors would like to thank I.~Gordeli, Y.~Oz and P.~Sodano for useful conversations and T.~Takayanagi for correspondence. DM gratefully acknowledges the hospitality of the Freiburg Institute for Advanced Studies, where the part of this work was done. The work was supported by the Brazilian Ministry of Science, Technology and Innovations and the graduate program of the DFTE-UFRN. The work of DM was also supported by the Science without Borders program of the CNPq-Brazil, MIT-IIP exchange program, by the Russian RFBR grant 14-02-00627 and the grant for support of Scientific Schools NSh~1500.2014.2.


\begin{appendix}

\section{Previous examples of AdS/BCFT}
\label{sec:examples}

In this appendix we review the solutions to~(\ref{Neumann-metric}) found in~\cite{taka} for the empty AdS case and for the $AdS_3$ black hole. The first case describes the asymptotic behavior of solution~(\ref{profileQ1}) and it is responsible for the divergence we discuss in section~\ref{sec:blackhole}. The second solution is a lower dimensional  analog of solution~(\ref{profileQ1}).

\subsection{Solution in empty $AdS$}
\label{sec:emptyAdS}

Let us recall how the AdS/BCFT construction works in the case of empty $AdS$ space, \emph{i.e.} the $f(z)=1$ case of metric~(\ref{ansatz-metric}). We are looking for the solution of the AdS/BCFT equations for the boundary CFT defined on a half-space $y<0$. Although we review the 4-dimensional case, $d+1$-dimensional generalization is straightforward~\cite{taka}.

In the situation with no extra matter on the boundary $Q$, equation~(\ref{Neumann-metric}) becomes
\be
\label{TakayanagiEq}
K_{ab}-(K-\Sigma)h_{ab}=0\,.
\ee
In empty $AdS$ space ($T=0$) this equation can be solved with tension $\Sigma\neq 0$.  Let us parameterize $Q$ as $y=y(z)$. The unit normal vector $n^\mu$ to $Q$ for a metric of the form~(\ref{ansatz-metric}) is given by the following expression.
\be
\label{normalQ}
(n^t,n^x,n^y,n^z)=\left(0,0,\,\frac{z}{L\sqrt{1+f(z)y'(z)^2}}\,,\,-\frac{z f(z)y'(z)}{L\sqrt{1+f(z)y'(z)^2}}\right).
\ee
The sign of the normal vector is fixed by the requirement that the vector is pointing outside of the space $N$ (figure~{\ref{fig:AdS/BCFT}}). The above sign corresponds to the situation $M: \{(t,x,y)|y\leq0\}$. In this convention the extrinsic curvature is computed as
 \be
 \label{ExtCurv}
 K_{\mu\nu}= {h_\mu}^{\alpha}{h_\nu}^{\beta}\nabla_\alpha n_\beta\,,
 \ee
 where ${h_\mu}^{\nu}$ is the projector onto $Q$:
 \be
 \label{projector}
 {h_\mu}^\nu = {\delta_\mu}^\nu - n_\mu n^\nu\,.
 \ee
The 3-dimensional tensors $K_{ab}$, $h_{ab}$ can be obtained via the projector
 \be
 h^{\mu}_a = \frac{\partial x^\mu}{\partial\tau^a}\,, \qquad e.g. \qquad K_{ab}= h^{\mu}_a h^{\nu}_b K_{\mu\nu} = h^{\mu}_a h^{\nu}_b \nabla_\mu n_\nu\,,
 \ee
where $\tau^a=\{t,x,z\}$ is a parametrization of $Q$.

One can readily find that~(\ref{TakayanagiEq}) is solved by
\be
\label{Qtension}
y'(z)=\, \frac{L\Sigma}{\sqrt{4-L^2\Sigma^2}}\,.
\ee
Introducing $L\Sigma=2\cos\theta$ one obtains
\be
\label{QprofileT}
Q\,:\quad y=z\cot\theta\,,
\ee
where $\theta$ is the angle between the positive direction of the $y$-axis and the hyperplane $Q$, or more generally, the tangential to $Q$ hyperplane at $z=0$, as shown on figure~\ref{fig:AdS/BCFT}.

Let us notice that the collection of half-hyperplanes~(\ref{QprofileT}) parameterized by the angle $\theta$ describes the foliation of the $AdS_4$ space in $AdS_3$ planes.  Indeed, the induced metric on $Q$ is just
\be
\d s^2 = \frac{L^2}{z^2}\left(-\d t^2 + \d x^2 +\frac{\d z^2}{\sin^2\theta}\right).
\ee
This way the embedding of $SO(2,2)$ into $SO(2,3)$ is realized geometrically.

For $T\neq 0$ there are no solutions to~(\ref{TakayanagiEq}) except for the tensionless brane $\Sigma=0$, which corresponds to
\be
\label{Qzerotension}
Q\,:\quad y={\rm const}\,, \qquad \text{or} \qquad \theta=\frac{\pi}{2}\,.
\ee
However, in the $d=2$ case black-hole solutions of the AdS/BCFT problem with $\Sigma\neq 0$ do exist. Let us now review this case.


\subsection{BTZ black hole}
\label{sec:BTZ}

\begin{figure}
 \centering
 \includegraphics[width=0.5\linewidth]{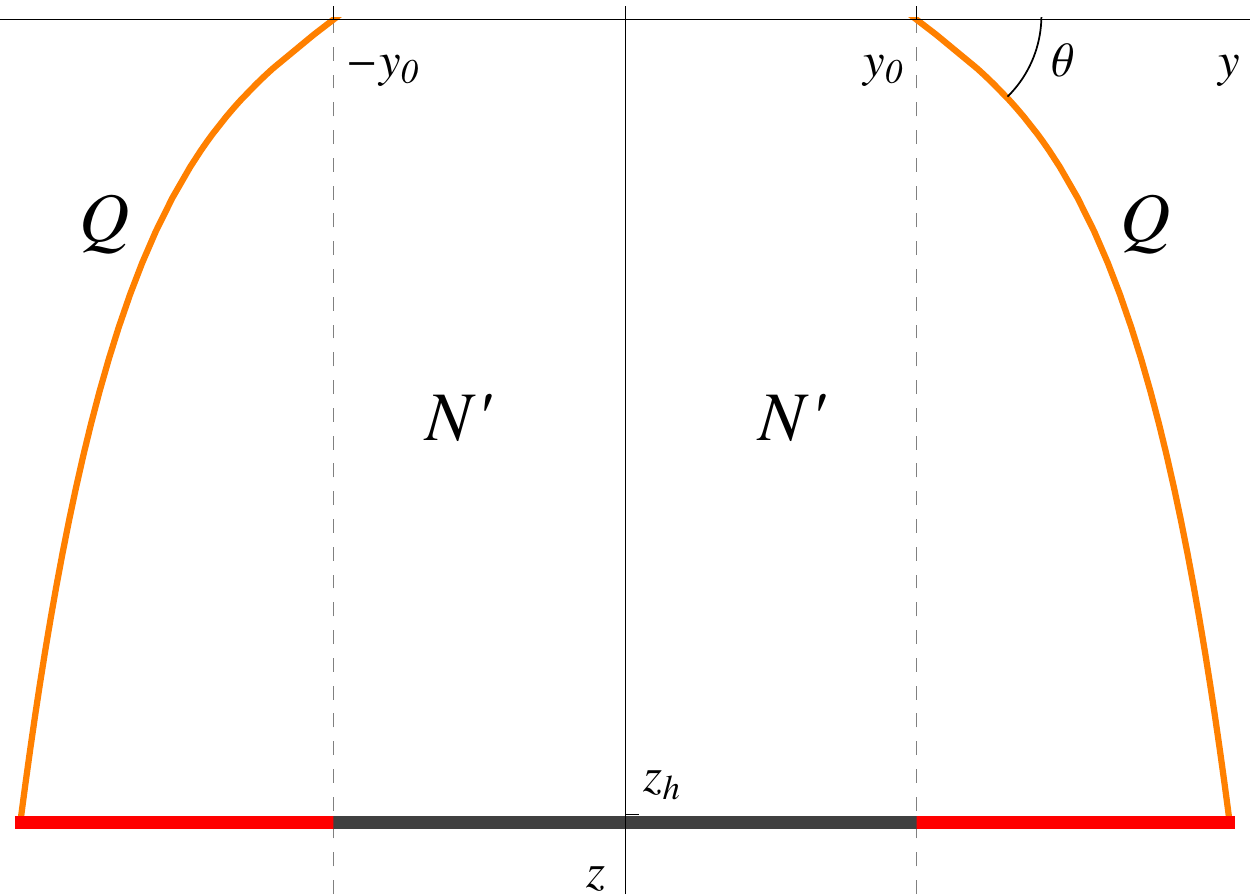}

 \caption{Profile of the boundary $Q$ in the case of the BTZ black hole.}
 \label{fig:BTZ}
\end{figure}

The problem of finding a boundary surface $Q$ for a black hole geometry was solved by Takayanagi in the $AdS_3/CFT_2$ case~\cite{taka}. The BTZ black hole~\cite{btz},
\be
\d s^2 = \frac{L^2}{z^2}\left(-f(z)\d t^2 + \d y^2 +\frac{\d z^2}{f(z)} \right), \qquad f(z)=1-\frac{z^2}{z_h^2}\,,
\ee
has a temperature given by
\be
T_{BTZ}=\frac{1}{2\pi z_h}\,.
\ee
The solution to equation~(\ref{TakayanagiEq}) for the profile $y(z)$ in this case is
\be
y(z)=y_0 + z_h\,{\rm arcsinh}\left(\frac{z}{z_h}\cot\theta'\right)\,, \qquad \cos\theta'=L\Sigma\,,
\ee
where $\theta'$ has the same meaning as $\theta$ in the previous section (figure~\ref{fig:BTZ}). For $z\to 0$, one reproduces the result of empty $AdS$~(\ref{QprofileT}). For $z\to z_h$ the profile enters the horizon at a finite angle, $y=y_0+\Delta y+\cos\theta'(z-z_h)$. Angle $\theta$ (tension), sets a characteristic distance scale $\Delta y=z_h{\rm arcsinh}(\cot\theta')$. It is natural to associate this scale with the ``width" of the boundary in the CFT.

We also notice that the profile of $Q$ is open in the direction of the horizon, that is $\theta\leq\pi/2$ for the right branch in figure~\ref{fig:BTZ} and $\theta\geq\pi/2$ for the left branch. This is a consequence of a ``null (weak) energy condition" on $Q$. One has to demand this in order for the temperature or energy density on $Q$ to be non-negative.

In~\cite{taka} the boundary entropy associated to this solution was computed. One first finds the full Euclidean action
\begin{multline}
I_{E}= -\frac{1}{2\kappa}\int_N \d^3x\, \sqrt{g_E}(R-2\Lambda) - \frac{1}{\kappa}\int_M \d^2x\, \sqrt{\gamma_E}(K^{(\gamma)}-\Sigma^{(\gamma)})  -\frac{1}{\kappa}\int_Q \d^2x\, \sqrt{h_E}(K-\Sigma)+\ldots=
\\ = I_{\rm bulk} + 2 I_{\rm bdry}+\ldots\,,
\end{multline}
where the ``bulk'' part, which does not depend on $\Sigma$, was separated from the $\Sigma$-dependent contribution of $Q$, and ellipses stand for the counter-terms necessary to remove the divergencies. The bulk part is given by
\begin{multline}
I_{\rm bulk}= -\frac{\beta}{2\kappa}\int\limits_{\epsilon}^{z_h}\d z\int\limits_{-\Delta y/2}^{\Delta y/2}\d y\,\sqrt{g_E}(R-2\Lambda)  - \frac{\beta}{\kappa}\int\limits_{-\Delta y/2}^{\Delta y/2}\d y\, \sqrt{\gamma_E}(K^{(\gamma)}-\Sigma^{(\gamma)})=
\\ = \frac{Lz_h}{2G}\Delta y\int\limits_{\epsilon}^{z_h}\frac{\d z}{z^3}  - \frac{Lz_h}{4G}\,\Delta y \,\frac{\sqrt{f(\epsilon)}}{\epsilon^2} = - \frac{L}{4G}\,\Delta y\left(\frac{1}{z_h}-\frac{z_h}{\epsilon^2}\right)- \frac{L}{4G}\,\Delta y \left(\frac{z_h}{\epsilon^2}-\frac{1}{2z_h}\right)+O(\epsilon)\,,
\end{multline}
where $\epsilon$ is an IR-regulator and $\Delta y=2y$ is now the length of the boundary interval. $\Sigma$-dependent part for one of the branches of $Q$ can be computed as follows.
\begin{eqnarray}
I_{\rm bdry} & = &  -\frac{\beta}{2\kappa}\int\limits_{\epsilon}^{z_h}\d z\int\limits_{y_0}^{y(z)}\d y\,\sqrt{g_E}(R-2\Lambda) - \frac{\beta}{\kappa}\int\limits_{\epsilon}^{z_h} \d z \,\sqrt{h_E}(K-\Sigma)= \nn
\\  & = &   \frac{Lz_h}{2G}\int\limits_{\epsilon}^{z_h}\frac{\d z}{z^3}\,y(z)-\frac{Lz_h}{4 G}\,\cos\theta'\,\int\limits_\epsilon^{z_h}\frac{\d z}{z^2}\, \frac{1}{\sqrt{1-f(z)\cos^2\theta'}} \nn
\\ & = & \frac{Lz_h}{4G\epsilon}\,\cot\theta' - \frac{L}{4G}\,{\rm arcsinh}({\cot\theta'}) + O(\epsilon)\,.
\label{Ibdry1}
\end{eqnarray}
Removing the divergent parts one arrives at the following expression for the entropy\footnote{The $O(\epsilon^{-2})$ divergence has been removed by the Gibbons-Hawking term at $M$, while one needs an extra counter-term to remove the $O(\epsilon^{-1})$ piece. We discuss the corresponding term in section~\ref{sec:blackhole} for the higher dimensional example.}
\be
\label{BTZentropy}
S = -\frac{\partial F}{\partial T}= \frac{L}{4G}\frac{\Delta y}{z_h}+ \frac{L}{2G}\,{\rm arcsinh}({\cot\theta'})\,.
\ee

We notice that the boundary contribution to the entropy is nothing but the area of the extra part of the horizon swept by the hypersurface $Q$. Strictly speaking this contribution must be corrected in the following way
\be
\label{BTZentropy-b}
S_Q =  \frac{L}{4G}\,{\rm arcsinh}(|\cot\theta'|)\,.
\ee
The non-analytic form of the entropy at $\theta'=\pi/2$ is related to the fact that going through the point $\theta'=\pi/2$ we should also change the direction of the normal to the surface $Q$ in order to satisfy the null (weak) energy condition. For the same reason the contribution of the second branch of $Q$ contributes the same amount to the entropy, rather than cancels the total result.


\section{Conformal fluids for constant $z$ slices}
\label{sec:conformalfluid}

In this appendix we review the simplest and most paradigmatic example of the fluid/gravity correspondence, which is given when considering $Q$ defined by the equation $z=z_0$ in the black hole geometry (\ref{ansatz-metric}). We choose the normal vector, which points towards the boundary of AdS space:
\be
\label{normalQz}
(n^t,n^x,n^y,n^z)=\left(0,0,0,-\,\frac{z \sqrt{f(z)}}{L}\right).
\ee
This choice is dictated by the requirement that the theory on $Q$ has positive energy density and temperature. This is then seen to be a degenerate limit of the AdS/BCFT where the part of the space with $z<z_0$ has been excised.

Here we only need the first three terms in (\ref{Tab}). $T_{ab}$ takes the form of the energy-momentum tensor of an ideal fluid
\be
\label{IdealFluid}
T_{ab}= (\epsilon+p)u_a u_b + p g_{ab}\,,
\ee
evaluated in the co-moving frame
\be
(u^t,u^x,u^y) = \left(\frac{z}{L\sqrt{f(z)}},0,0\right).
\ee
The energy density and the pressure are thus given by
\begin{eqnarray}
\label{ezconst}
\epsilon & = &  u^a u^a T_{ab} = \frac{L^{2}}{\kappa z^3}\left(L\Sigma-2\sqrt{f}\right)  \\
\label{pzconst}
p & = & \frac{\epsilon + {T^a}_a}{2}= -\frac{L^2 \left(z f'+2 L \Sigma \sqrt{f}-4 f\right)}{2\kappa z^3 \sqrt{f(z)}}\,,
\end{eqnarray}
and we get the following equation of state
\be
\frac{\epsilon}{p}=\frac{4f-2L\Sigma\sqrt{f}}{zf'-4f+2L\Sigma\sqrt{f}}\,.
\ee
In the empty $AdS$-space, $f(z)=1$, and general $\Sigma$, $T_{ab}$ describes an energy dominated universe. The hydrodynamical quantities diverge as $z\to 0$. For a finite temperature case this is also the leading order result for general $\Sigma$ in the $z\to 0$ asymptotics. The choice $\Sigma=2/L$ removes the UV divergences and is called holographic renormalization. In fluid/gravity correspondence this means subtracting non-hydrodynamic degrees of freedom from the energy-momentum tensor. The resulting equation of state at the $AdS$ boundary is that of a conformal incompressible fluid
\be
\epsilon = 2 p\,, \qquad \text{with} \qquad \epsilon= \frac{L^2}{\kappa z_h^3}
\ee

One can also find the local temperature and local entropy density of the thermodynamical system~(\ref{ezconst})-(\ref{pzconst}). The local temperature $T$ should be given by the red-shifted Hawking temperature $T_H$~(\ref{HawkingT})
\be
T=\frac{T_H}{\sqrt{f(z)}}\,.
\ee
From the thermodynamic relation
\be
\label{ensemble}
\epsilon +p = Ts\,,
\ee
one can find that the entropy density is precisely the Bekenstein-Hawking entropy density
\be
s=\frac{L^2}{4Gz_h^2}\,.
\ee
The consistency of the first law of thermodynamics can be verified locally through the relations
\be
\label{1stlaw}
\frac{ds}{s}=\frac{d\epsilon}{\epsilon+p}\,, \qquad \frac{dT}{T}=\frac{dp}{\epsilon + p}\,.
\ee

These results are well-known in the study of fluid-gravity correspondence, \emph{e.g.} see reviews~\cite{Rangamani:2009xk,Hubeny:2010wp,f/g-review} and references therein. Changing the position of the surface along the $z$ axis one can study the renormalization of the physical quantities in the hydrodynamical theory~\cite{Brattan:2011my}. It is noticed that the speed of sound, see figure~\ref{fig:density-pressure} (right), given by
\be
v_s^2 = \frac{\partial p}{\partial\epsilon}=\frac{\partial p/\partial z_h}{\partial\epsilon/\partial z_h}= \frac{3-f(z)}{4f(z)}\,,
\ee
becomes superluminal for $z>z_s$,
\be
z_s^3 = \frac25\,z_h^3\,.
\ee
In~\cite{Brattan:2011my} it was suggested that this indicates that the gravity solution must be modified beyond this point.

\section{Useful formulae}
\label{sec:appendix}
\begin{eqnarray}
\int\limits_0^1\frac{\d \zeta}{\sqrt{1-\zeta^3}} & = & \frac{\sqrt{\pi }\, \Gamma \left(\frac{4}{3}\right)}{\Gamma \left(\frac{5}{6}\right)}\,, \\
\int\limits_0^1\d \zeta\,\frac{1-\sqrt{1-\zeta^3}}{\zeta^3\sqrt{1-\zeta^3}} & = & \frac12 + \frac14 \int\limits_0^1\frac{\d \zeta}{\sqrt{1-\zeta^3}}\,, \\
\int\limits_0^1\d \zeta\,\frac{\sqrt{1-\zeta^3}-1}{\zeta^3} & = & \frac12 - \frac34 \int\limits_0^1\frac{\d \zeta}{\sqrt{1-\zeta^3}}\,.
\end{eqnarray}

\end{appendix}

\addcontentsline{toc}{section}{Bibliography}

\bibliography{biblioboundary}
\bibliographystyle{ieeetr}

\end{document}